\documentclass{emulateapj}

\newcommand{\swift}{{\it Swift}}

\newcommand{\rosat}{{\it ROSAT}}

\newcommand{\xmms}{{\it XMM-Newton }}
\newcommand{\chandra}{{\it Chandra}}
\newcommand{\ax}{$\alpha_{\rm X}$}

\newcommand{\plm}{$\pm$}

\newcommand{\lledd}{$L/L_{\rm Edd}$}

\def\eps@scaling{1.0}%

\newcommand\plotthreetwo[2]{{%
 \typeout{Plotthree included the files #1 }
 \centering
 \leavevmode
 \columnwidth=.40\columnwidth
 \includegraphics[width={\eps@scaling\columnwidth}]{#1}%
 \hfill
 \includegraphics[width={\eps@scaling\columnwidth}]{#2}%
}}%

\shorttitle{A new outburst of IC 3599}
\shortauthors{Grupe et al.}

\begin{document}

\title{IC 3599 did it again: A second outburst of the X-ray transient Seyfert 1.9 Galaxy 
}

\author{Dirk Grupe\altaffilmark{1},
S. Komossa\altaffilmark{2},
Richard Saxton\altaffilmark{3}
}

\altaffiltext{1}{Department of Earth and Space Science, 
Morehead State University,
235 Martindale Dr., Morehead, KY 40351, U.S.A.; d.grupe@moreheadstate.edu} 

\altaffiltext{2}{Max-Planck-Institut f\"ur Radioastronomie, Auf dem H\"ugel 69,
53121 Bonn, Germany
}

\altaffiltext{3}{European Space Astronomy Centre (ESAC), 
P.O. Box, 78,  E-28691 Villanueva de la Ca\~nada, Madrid, Spain
}

\begin{abstract}

We report on the \swift\ discovery of a second high-amplitude (factor 100) outburst 
of the Seyfert 1.9 galaxy IC 3599, and discuss implications
for outburst scenarios. 
\swift\ detected this active galactic nucleus (AGN) again in February 2010
in X-rays at a level of 
(1.50\plm0.11)$\times 10^{36}$ W (0.2-2.0 keV), which is
nearly as luminous as the first outburst detected with \rosat\ in 1990.
Optical data from the Catalina sky survey show that the optical emission was already
bright two years before the \swift\ X-ray high-state. 
Our new \swift\ observations
performed between 2013 and 2015
show that IC 3599 is currently
again in a very low X-ray flux state.
This repeat optical and X-ray outburst, and the long optical duration,
suggest that IC 3599 is likely
not a tidal disruption event (TDE). Instead, variants of AGN-related variability
are explored. The data are consistent with an accretion disk instability
around a black hole of mass on the order 10$^6$--10$^7$ M$_{\odot}$; a value
estimated using several different methods.

\end{abstract}


\section{Introduction}
The highest amplitudes of X-ray variability recorded so far among galaxies have been detected from
the cores of a few quiescent galaxies, and have been interpreted as the tidal disruption and
accretion of stars by dormant supermassive black holes \citep[SMBHs; review by][]{komossa12}.
These events happen when a star gets too close to the
central SMBH and is disrupted due
to tidal forces. Part of the stellar 
debris will form an accretion disk and will be swallowed by the SMBH \citep[e.g.,][]{rees88},
 producing a luminous accretion flare.   One of the most
spectacular of these events, Swift J1644+57, was discovered in March 2011 by \swift\
\citep{bloom11, burrows11}.
Before \swift\ \citep{gehrels04}, one of the most successful surveys to discover TDEs 
was the \rosat\
All-Sky Survey (RASS), during which several events were found
 \citep[]{bade96, grupe99, komossa99b, komossa99a}.

The Seyfert 1.9 galaxy IC 3599 
\citep[Zw 159.034; 1RXS J123741.2$+264227$;
$\alpha_{2000}$ = $12^{\rm h} 37^{\rm m} 41.^{\rm s}2$, 
$\delta_{2000}$ = $+26^{\circ} 42' 27^{''}$, z=0.0215;][]{grupe95, brandt95,
komossa99a}
was noticed in the RASS due to its very bright X-ray emission. 
However, when \rosat\ re-observed IC 3599 again
about a year later its flux (0.1-2.4 keV) had 
dropped by a factor of more than 60 \citep{grupe95, brandt95}, 
and continued to fade, implying a total amplitude
of variability larger than a factor 100.
Its X-ray spectrum was very soft, and the high-state X-ray emission
was accompanied by strong, broad emission lines from hydrogen, helium,
and iron \citep{brandt95},
 which faded significantly in subsequent years \citep{grupe95, komossa99a}.
This kind of emission-line flaring is very rare, and has only recently been
observed in a few other galaxies \citep[e.g.,][]{koetal08}.
A broad H$\alpha$ line component
was still present 7 years after the high-state of IC\,3599 (and still is in a
recent SDSS spectrum; see below), which led to a Seyfert 1.9 classification
of the galaxy \citep{komossa99a}.  
The classification of IC 3599 as an AGN is consistent with its radio 
detection \citep{bower13}
and with its AGN-dominated MIR spectrum \citep{sani10}.
The cause for the outburst of IC 3599 remained unknown, and several possibilities
were considered: high-amplitude Narrow-line Seyfert 1 (NLS1) 
variability{\footnote{Note, however,
that IC 3599 is very likely not a NLS1 galaxy, since it does not show
any FeII emission, neither at outburst nor quiescence, and since its Seyfert 1.9
classification indicates that the inner, high-velocity BLR is obscured}}, 
an accretion disk instability, or a TDE. 
Its bright high-ionization iron lines were successfully modeled by photoionization
of high-density gas with the properties of a CLR (coronal line region), illuminated
by a flare with a strong EUV -- soft X-ray bump \citep{komossa99a}. The faint 
coronal lines still present years after the outburst did not vary further, indicating
that they are permanent, and similar to those seen in other Seyfert galaxies. 

When \swift\ observed IC 3599 for the first time in 2010 February, it
detected a new outburst. The 
luminosity level of 1.5$\times 10^{36}$ W 
(0.2-2.0 keV band)
was only a factor of 2 below the value seen during the RASS 
\citep{grupe95, grupe01}.
After we noticed the high-state, we triggered new \swift\ follow-ups of
IC 3599 which are reported here, along with other multi-wavelength data
and a re-discussion of outburst scenarios. 
Results of our study were first presented by \citet{komossa15},
and independently by \citet{campana15}, who explored the possibility 
that IC 3599 is experiencing repeat tidal stripping of an orbiting star.  

Throughout this paper spectral indices are denoted as 
$F_{\nu} \propto \nu^{-\alpha}$. Luminosities are calculated assuming a $\Lambda$CDM
cosmology with $\Omega_{\rm M}$=0.286, $\Omega_{\Lambda}$=0.714 and a Hubble
constant of $H_0$=70 km s$^{-1}$ Mpc$^{-1}$. This results in a luminosity
 distances $D$=91.4 Mpc
using the cosmology calculator  by \citet{wright06}.
 All errors are 1$\sigma$ unless stated otherwise.

\section{\label{observe} Observations and Data Reduction}

\subsection{\swift\ Data}

A full list of the \swift\ observations of IC 3599 is given 
in Table\,\ref{swift_res}. 
The \swift\ X-ray telescope (XRT)
was operating in photon counting mode  and the
data were reduced by the task {\it xrtpipeline} version 0.12.6., 
which is included in the HEASOFT package 6.12. Source counts were selected in a
circle with a radius of 70.7$^{''}$ for the 2010 February and May observations, and 
47.1$^{''}$ for all others. 
{\bf The larger source extraction radius during the 2010 observations was 
chosen because the source was brighter.
}
The
background counts were collected in a nearby 
circular region with a radius of 247.5$^{''}$. 
Only the 2010 February and May data had enough counts to allow for 
a spectral analysis of each individual observation. 
The data of 2013 and 2014 
were merged to allow for a spectral analysis. 
For all spectra we used the most recent response file {\it swxpc0to12s6\_20130101v014.rmf},
and {\it XSPEC} version 12.7.1 for analysis.
Due to the small number of counts used in the spectra, the counts were not binned.

The UV-optical telescope (UVOT)
data of each segment were coadded in each filter with the UVOT
task {\it uvotimsum}. Except for 
2013 November, all observations were performed in all 6 UVOT filters.
Source counts in all 6 UVOT filters
  were selected in a circle with a radius of 3$^{''}$ and background counts in
  an annulus around the source extraction region, with radii of 
  $R_{\rm in}=5^{''}$ and $R_{\rm out}=20^{''}$, in order to subtract the contribution
  of star light from the host galaxy. 
  UVOT magnitudes and fluxes were measured with the task {\it  
uvotsource} and aperture corrected by setting the parameter 
{\it apercorr=CURVEOFGROWTH}
based on the most recent UVOT calibration as described in  
\citet{breeveld10}.
The UVOT data were corrected for Galactic reddening
\citep[$E_{\rm B-V}=0.015$;][]{schlegel98}, using 
equation (2) in \citet{roming09}.

\subsection{\xmms Slew Survey}

The field of IC 3599 was covered during
the \xmms slew  9137800003 \citep{saxton08},
on 2007-June-19. No photons from the source were detected and
the EPIC pn, 2-sigma, upper limits of 0.8 and 0.7 counts s-1, imply
flux upper limits of 1.2$\times 10^{-15}$ 
and 9.6$\times 10^{-16}$ W m$^2$ in the
0.2-12 keV and 0.2-2 keV bands respectively, assuming a spectrum with 
slope \ax=2.5 and Galactic 
absorption of $N_{\rm H}=1.16 \times 10^{20}$ cm$^{-2}$.

\section{\label{results} Results}

\subsection{X-ray Spectral Analysis}
The X-ray spectrum during the outburst in 2010 February can be fitted with a simple absorbed power law
model with a X-ray spectral index \ax=2.63\plm0.30 and the absorption column density fixed to the Galactic value.
 The flux in the 0.2-2.0 keV band assuming this
model is (1.41\plm0.11)$\times 10^{-14}$ W m$^{-2}$. This is about half the flux that was detected during
the RASS \citep{grupe95,  grupe01}. 
The 0.3-10 keV flux was 9.26$^{+0.56}_{-0.54} \times 10^{-15}$ W
m$^{-2}$. The spectral slope is similar to that measured during the RASS.
Alternatively, the X-ray spectrum can also be fitted by an absorbed blackbody
plus power law model (with \ax\ fixed to 1.0) which results in a blackbody
temperature equivalent of $kT$ = 91\plm16 eV. This is the same value as measured
during the original outburst during the RASS \citep{grupe95}.

The Seyfert 1.9 classification of IC 3599 does suggest that we should expect to see excess
absorption above the Galactic value. Although there is some hint of excess absorption
when the spectrum is fitted by a single power law model, there is no evidence for
excess absorption when the data are
fitted by a blackbody model. However, the quality of the
relatively short-exposed spectrum during the high state in February 2010 does not
allow final conclusions on the intrinsic column density.

In order to obtain an X-ray low state spectrum, we combined the data from the \swift\ observations of 2013
and 2014 which resulted in a total exposure time of 16.9 ks with
a total of 50 source counts. The spectral analysis was performed applying Cash
statistics \citep{cash79}. This spectrum can be modeled with a harder spectrum with an X-ray
spectral index of \ax=1.40\plm0.70 
{\bf with $N_{\rm H}$ fixed to the Galactic value. 
There is no indication of excess absorption beyond that.}
The 0.3-10 keV flux of this combined spectrum 
is (9.67\plm1.62)$\times 10^{-17}$ W m$^{-2}$ -- a factor 100 lower 
than during the 2010 February outburst. 
Note that when applying the absorbed blackbody plus power law model mentioned
above during the high state, the black body component has disappeared in the low
state spectrum.

\subsection{Long-term Light Curve}

\begin{figure}
\epsscale{0.5}
\plotone{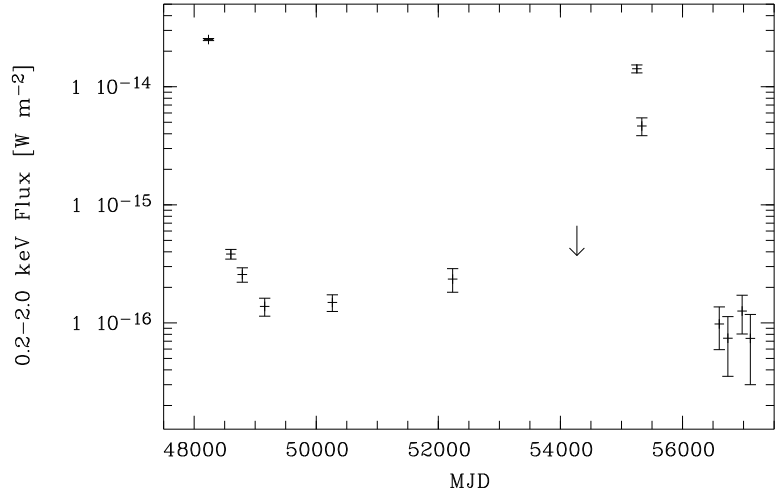}
\caption{Long-term 0.2-2.0 keV light curve of IC 3599, starting with the RASS observation.
 \label{swift_long_lc}
}
\end{figure}

The fluxes in X-rays and in the UV/optical obtained by \swift\ 
are listed in Table\,\ref{swift_res}.  
The \rosat\ observations have already been
 published in \citet{grupe95, grupe01} and  \citet{brandt95},
 and the \chandra\
 measurements were discussed in \citet{vaughan04}. The long-term light curve
 including our latest \swift\ observations is shown in the left panel of
 Figure\,\ref{swift_long_lc}
 \citep[see also][]{komossa15, campana15}, 
 together with the \xmms slew-survey upper limit.

\subsection{Catalina Sky Survey Optical Light Curve}

\begin{figure}
\epsscale{0.6}
\plotone{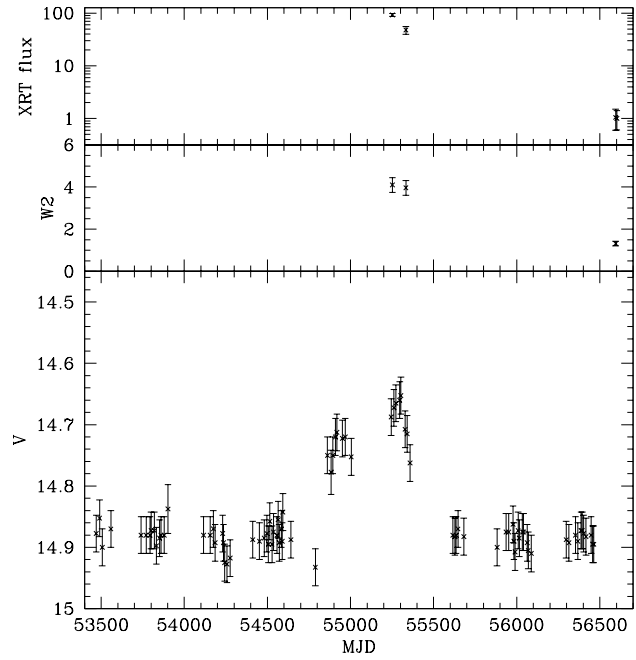}
\caption{
 \swift\ XRT and W2 light curves of IC 3599 and in the lower 
 panel the optical light curve obtained by the Catalina Sky Survey. 
XRT flux is in units of 10$^{-16}$ W m$^{-2}$
and W2 flux is in units of 10$^{-15}$ W m$^{-2}$. 
 \label{swift_catalina_lc}
}
\end{figure}

While we do not have X-ray observations between the \chandra\ and \xmms slew
survey observations in 2002 and 2007, respectively, IC 3599 was monitored in
the optical by the Catalina sky survey \citep{drake09}.
The Catalina light curve  is shown together with the
\swift\ XRT and UVOT W2 light curves in the right panel of
 Figure\,\ref{swift_catalina_lc}. According to the Catalina light curve, 
IC 3599 was already optically bright in 2008, and 
the outburst must have happened around MJD 54570 which translates to
2008 April 14. 
Then, either a second optical outburst occurred in 2010, or else 
the outburst lasted for roughly two years with the
peak at around MJD 55260 (2010 March 05) which was about a week after the 
\swift\ observation on 2010 February 25. 
These observations provide potentially tight constraints on outburst scenarios.

\subsection{Spectral Energy Distributions}

\begin{figure}
\epsscale{0.5}
\plotone{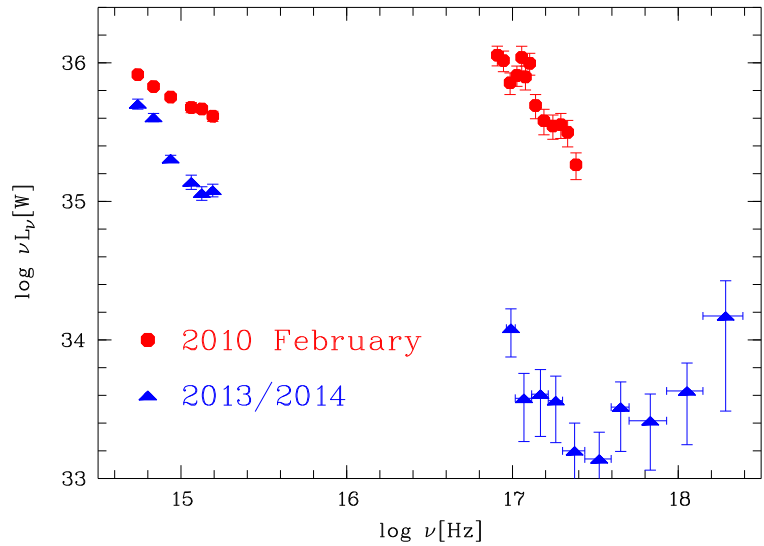}
\caption{Development of the spectral energy distribution between the 2010
 February high state and the 2013/2014 low state.
 \label{ic3599_sed}
}
\end{figure}

The spectral energy distributions during the February 2010 high state 
and the 2013/2014 low state are
displayed in 
Figure\,\ref{ic3599_sed}. 
The drop seen in X-rays is significantly larger 
than in the optical/UV  (note, however, that the optical emission 
may still be affected by host galaxy contribution. 
If so, these values are upper limits for the core emission).  
Under these constraints,
and assuming a black hole mass of $10^7 M_{\odot}$ (see below), the
Eddington ratios during these two epochs were \lledd=0.04 and 0.003,
respectively.

\subsection{The Optical SDSS Spectrum}

A new optical spectrum of IC 3599 was taken in the course of the Sloan Digital Sky
Survey \citep[SDSS,][]{york00}
in 2005-December-25 (MJD 53729). 
We have used this spectrum for a SMBH mass estimate, and in order to check, 
whether emission lines have flared again, as previously observed \citep{brandt95}.
We find that narrow emission lines including [FeVII] transitions are at their low emission
levels, as seen in previous low-state optical spectra from 1995 \citep{grupe95}
and 1997 \citep{komossa99a},
 and emission-line ratios are consistent with little or
no variability since then. 
An analysis of the H$\alpha$ emission line complex shows, that the spectrum is still
consistent with the presence of a faint broad emission line component, 
with FWHM(H$\alpha$)$\sim 1020$ km s$^{-1}$, 
similar to the value measured before 
(Brandt et al. 1995, Grupe et al. 1995, Komossa \& Bade 1999).  
Because of the Seyfert 1.9 classification, we do not use this line width to estimate
the SMBH mass of IC 3599, since application of the well-established width -- 
luminosity relation \citep[e.g. ][]{peterson04}
requires an unobscured view on the bulk of the broad-line region (BLR). 
Instead, we use the width of [OIII]$\lambda$5007 as a proxy for velocity dispersion 
\citep[e.g.][]{nelson00},
 and then use the scaling relation with SMBH mass as given by \citet{tremaine02}.
 First, we have fit the [OIII] emission with a Gauss profile, and 
have carefully 
checked that [OIII] is well represented by a single component, and does not show a 
strong asymmetry, or an extra blue wing.  We then obtain a line width 
of FWHM([OIII]) = 280 km/s (corrected for instrumental resolution), well 
consistent with
the value of 260 km/s reported by \citet{komossa99a}.
 Applying the relation of \citet{tremaine02}
 then gives $M_{\rm BH}$ = 1.2 $\times10^7$ M$_{\odot}$.

\section{\label{discuss} Discussion}

We have detected, and followed up,
a second outburst in the Seyfert galaxy IC 3599 with \swift. 
This has important implications for outburst scenarios of this unusual source,
 because
some of those initially considered, did not predict repeat flaring. 
Since the value is important for the further discussion, we
first comment on several estimates of the SMBH mass of IC 3599, and then proceed
with discussing outburst scenarios.

\subsection{Black Hole Mass Determination}

We have followed several different approaches of estimating the SMBH mass
of IC\,3599.
First, applying scaling relations between velocity dispersion
 $\sigma_{\rm [OIII]}$
and SMBH mass, we obtain $M_{\rm BH}$ = 1.2 $\times 10^7 M_{\odot}$ (Sect. 3.5).
Second, from the relationship of black hole mass to bulge K-band luminosity
\citep{Marconi03} we find $M_{BH}\sim2.2\times10^{6}M_{\odot}$, with
a systematic error of 0.3 dex, from the 2MASX magnitude of $k=11.97$,
after subtracting 25\% contributed by the central AGN-dominated  point source and
correcting for the bulge ratio of 0.42 \citep{simard11}.
Third, an order of magnitude estimate of the SMBH mass 
can also be derived from the temperature of the accretion
disk measured from the high-state spectrum in 2010, $kT=91$ eV. 
 Following \citet[][equation 3.20]{peterson97}, we obtain
$\sim 2.5\times 10^6 M_{\odot}$.

All three estimates are in the range (2--12)$\times 10^6$ M$_{\odot}$.
We note in passing, that this is much higher than the value used by \citet{campana15},
$3\times 10^5 M_{\odot}$, who used the width of {\em narrow} 
H$\beta$ of \citet{sani10}
and applied BLR scaling relations. 

\subsection{Outburst Scenarios} 

\subsubsection{Repeat Tidal Disruption Events}

At first glance, the detection of
repeated flaring is unexpected, if the TDE interpretation of IC 3599 was
correct, because classical TDEs are rare events, at a rate of about 
10$^{-4}$--10$^{-5}$ per year and per galaxy 
\citep[][for reviews of theory and observations, respectively]{alexander12,
komossa12}.
However, recent work has shown that recurrent outbursts or high states may also occur in TDE scenarios,
for instance when one of the following conditions is met: 
(1) If IC\,3599 hosts
a binary black hole or recoiling black hole, tidal disruption rates are temporarily
strongly boosted \citep[e.g.][]{chen09, komossa08, stone11}, and a new disruption
event can occur within decades. 
(2) If a TDE in IC\,3599 happened in a  binary SMBH,
its light curve would show characteristic
recurrent dips, since the presence of the secondary temporarily
interrupts the accretion stream on the primary \citep{liu14}.
However, in both cases it is unlikely that they produce a peak luminosity
very similar to the earlier one in 1990. 
(3) A third possibility was studied by \citet{campana15}, who suggested that
 the flaring
is due to repeat tidal stripping of an orbiting star. However, this scenario, 
which predicts
a rapid rise to maximum (their Fig. 2) may be difficult to reconcile with the 
optical high-state already seen in 2008.

\subsubsection{Variability Related to a Supermassive Binary Black Hole} 

If the outbursts of IC 3599 continue repeating, 
the behavior is reminiscent of the blazar OJ 287 \citep[e.g.][]{sill88, valtonen08},
 and a binary
SMBH  might be responsible; i.e., a secondary BH impacting
the accretion disk around the primary while orbiting.
The two optical high-states in 2008 and 2010 are then similar to the double-peaked 
optical high-states of OJ 287, which have been interpreted as the two disk impacts
of the secondary BH during its orbit.  
Another mechanism which produces repeat outbursts is episodic stream-feeding of one of the
black holes in a binary SMBH system \citep{tanaka13} which recurs every orbit.
Ongoing monitoring with \textit{Swift} is required in order to test these 
scenarios further.

\subsubsection{A Disc Instability}
Given the evidence that IC\,3599 hosts a long-lived
AGN, extreme processes in its accretion disk are a likely cause of the repeat outbursts. 
These are suggestive of a recurrent regular process.
Similar large flares have been seen in a small number of accreting Galactic sources 
\citep[e.g. GRS~1915+105, GRO~J1744-28, and IGR~17091-3624;][respectively]
{belloni97, cannizzo96, pahari13}. In these cases the
 fast rise and decay in flux has been attributed to an instability in a radiation-pressure dominated
 inner disk \citep{lightman1974}. Recently, \citet{saxton15}
  have applied the
instability to a single flare seen in the nucleus of the galaxy NGC~3599. In this 
mechanism, the inner disc is initially empty or filled with tenuous gas out to a 
truncation radius, $R_{trunc}$. During the quiescent phase the inner disc is filled by 
diffusion from the outer disc without significantly increasing the X-ray flux.
 At some point, the internal radiation pressure of the disc becomes greater than 
 the gas pressure 
causing a heating wave, that increases the local viscosity, scale height and accretion
rate, to propagate from the inner radius back to the truncation radius. In this phase
the luminosity increases quickly. Later, as the
inner disc is accreted more rapidly than it is replenished, the flare decays back
to the quiescent level, a new hole is left in the central region and 
the cycle begins again. 

The repeat time for flares is given by the filling time of the inner disc 
\citep{saxton15} as 

\begin{equation}
    \tau_{fill} \sim 0.33\alpha^{-8/10}M_{6}^{6/5}M_{edd}^{-3/10} \left[ 
    \left(\frac{R_{trunc}}{R_{g}}\right)^{5/4} -
    \left(\frac{R_{0}}{R_{g}}\right)^{5/4} \right] 
\end{equation}

in units of months
where  $M_{6}$ is
the black hole mass in units of $10^{6}$ M$_{\odot}$ and $R_{g}=GM/c^{2}$, 
the gravitational radius.
Assuming that the infill rate is Eddington limited ($M_{edd}=1$), 
the viscosity $\alpha=0.1$ and the inner radius $R_{0}=3R_{g}$ then for a flare 
repeat time of 
19.5 years and $M_{BH}=10^{6} - 10^{7} M_{\odot}$ 
we find 
$R_{trunc}$=5--45 $R_{g}$. This is comparable to the value of
4--22 $R_{g}$ (depending on spin) which was inferred in GRS~1915+105 
\citep{belloni97, belloni97a}.

A lower limit for the rise time is given by
the time taken to fully heat the inner disc \citep{belloni97, saxton15}:

\begin{equation}
    \tau _{rise} \gtrsim 1.5\times10^{4} \left( \frac{R_{trunc}}{R_{g}}\right) M_{6}\,\mathrm{seconds} 
\end{equation}

For the flare in IC 3599, $\tau_{rise}\gtrsim7-8$ days,
{\bf consistent with the $\tau_{rise}\leq2.4-17$ months
inferred from the Catalina curve.} \footnote{Several mechamisms which could extent the flare duration have been 
explored \citep{nayakshin00, janiuk05}}
This scenario is supported by the X-ray spectra which can be explained by
a weak power-law component
augmented by a highly variable thermal component.

\acknowledgments
We thank our anonymous referee for
valuable comments and suggestions, and 
Neil Gehrels for approving our various \swift\ ToO requests and
the \swift\ science operation team for performing the observations. 


\begin{deluxetable}{ccclrrrrrr}
\tablecaption{\swift\ XRT and UVOT fluxes of IC 3599 \label{swift_res}
}
\tablewidth{0pt}
\tablehead{
\colhead{$T_{\rm start}$\tablenotemark{1}} &
\colhead{MJD} & 
\colhead{$\rm F_{\rm 0.3-10 keV}$\tablenotemark{2}} &
\colhead{$\rm F_{\rm V}$\tablenotemark{3}} &
\colhead{$\rm F_{\rm B}$\tablenotemark{3}} &
\colhead{$\rm F_{\rm U}$\tablenotemark{3}} &
\colhead{$\rm F_{\rm W1}$\tablenotemark{3}} &
\colhead{$\rm F_{\rm M2}$\tablenotemark{3}} &
\colhead{$\rm F_{\rm W2}$\tablenotemark{3}} 
} 
\startdata
2010-02-25 07:46 & 55252.365 & (9.26$^{+0.54}_{-0.56})\times 10^{-15}$       
  & 8.19\plm0.40 & 6.72\plm0.33 & 5.64\plm0.34 & 4.73\plm0.40 & 4.63\plm0.28 & 4.10\plm0.35 \\
2010-05-17 00:58  & 55333.049 & (4.77\plm0.82)$\times 10^{-15}$      
  & 7.39\plm0.46 & 6.11\plm0.33 & 5.01\plm0.37 & 4.17\plm0.43 & 4.39\plm0.31 & 3.95\plm0.35 \\
2013-10-30 00:34  & 56595.160 & (1.05$^{+0.50}_{-0.41}) \times 10^{-16}$
&  4.67\plm0.27 & 3.85\plm0.20 & 2.04\plm0.15 & 1.54\plm0.14 & 1.26\plm0.10 & 1.31\plm0.11 \\
2013-11-06 02:24  & 56602.208 & (1.01$^{+0.41}_{-0.34}) \times 10^{-16}$
& \nodata & \nodata & \nodata & \nodata & 1.26\plm0.07 & \nodata \\
2014-03-26 07:50  & 56742.583 & (7.12$^{+3.54}_{-2.75}) \times 10^{-17}$
&  4.94\plm0.39 & 4.11\plm0.27 & 2.24\plm0.20 & 1.56\plm0.18 & 1.44\plm0.09 & 1.36\plm0.14 \\
2014-08-08 07:07  & 56877.306 & $<2.9\times 10^{-16}$
      & 4.91\plm0.37 & 4.00\plm0.25 & 2.21\plm0.19 & 1.73\plm0.18 & 1.47\plm0.14 & 1.33\plm0.13 \\
2014-11-15 04:12  & 56976.188 & (1.51$^{+0.42}_{-0.36}) \times 10^{-16}$
      & 5.03\plm0.42 & 4.02\plm0.28 & 2.03\plm0.12 & 1.38\plm0.16 & 1.14\plm0.13 & 1.20\plm0.13 \\
2014-11-23 08:57  & 56984.375 & $<6.9\times 10^{-16}$
      & \nodata & \nodata & 2.10\plm0.15 &  \nodata & \nodata & \nodata \\
2015-03-29 21:37 & 57110.969 & (4.7$^{+3.9}_{-2.2})\times 10^{-17}$ & 
       4.84\plm0.28 & 3.75\plm0.21 & 2.02\plm0.15 & 1.32\plm0.13 & 1.20\plm0.10 & 1.08\plm0.10
\enddata

\tablenotetext{1}{Start times are given in UT}
\tablenotetext{2}{Observed 0.3-10 keV flux in units of W m$^{-2}$. Fluxes from 
the observations of 2013 and 2015 were based on an energy conversion factor 
derived from the 2010-05-17 data and the uncertainties were
determined using the method described in \cite{kraft91}. Upper limits are 
3$\sigma$ as described in \cite{kraft91}. 
}
\tablenotetext{3}{The reddening corrected 
fluxes in the UVOT filters are given in units of $10^{-15}$ W m$^{-2}$.}
\end{deluxetable}

\end{document}